\documentclass[%
twocolumn,
superscriptaddress, 
 amsmath,amssymb,
 aps,
 prl,
 floatfix
]{revtex4-2}

\usepackage{graphicx}
\usepackage{dcolumn}
\usepackage{bm}
\usepackage{xr-hyper}
\usepackage{hyperref}
\usepackage{xcolor}
\hypersetup{
    colorlinks=true,
    citecolor=blue,
    linkcolor=blue,
    urlcolor=blue,
}

\begin{document}


\preprint{APS/123-QED}

\title{Avalanche criticality emerges by thermal fluctuation in a quiescent glass}

\author{Yuki Takaha}
 \email{yukitakaha@g.ecc.u-tokyo.ac.jp}
 \affiliation{Graduate School of Arts and Sciences, The University of Tokyo, 3-8-1 Komaba, Tokyo 153-8902, Japan}
\author{Hideyuki Mizuno}
 \affiliation{Graduate School of Arts and Sciences, The University of Tokyo, 3-8-1 Komaba, Tokyo 153-8902, Japan}
\author{Atsushi Ikeda}
 \affiliation{Graduate School of Arts and Sciences, The University of Tokyo, 3-8-1 Komaba, Tokyo 153-8902, Japan}
 \affiliation{Research Center for Complex Systems Biology, Universal Biology Institute, The University of Tokyo, Komaba, Tokyo 153-8902, Japan}

\date{\today}

\begin{abstract}
We report avalanche criticality of thermal relaxation in glassy systems after a rapid quench by molecular simulation.
Our analysis of the energy landscape and the scaling reveals that particle rearrangement is critical.
The critical phenomenon has the same origin as avalanches in sheared amorphous solids, but the critical exponent differs from previously observed.
Our results suggest that by viewing a glass as a thermally driven elastoplastic material, we can understand dynamics below the glass transition point, such as aging.
\end{abstract}

\maketitle


\textit{Introduction}.---Various systems respond to changing external conditions with intermittent events, so-called avalanches, of various sizes, which follow a power-law distribution.
Earthquakes following the Gutenberg--Richter rule~\cite{fisher1997statistics, dahmen1998gutenberg}, crackling noise in crumpled sheets~\cite{kramer1996universal, shohat2023logarithmic}, response of random magnetic materials to magnetic fields~\cite{sethna1993hysteresis}, and fluidization of amorphous solids~\cite{maloney2006amorphous} are examples of such phenomena.
Although these systems consist of completely different components and have different rules of microscopic dynamics, a scale-free behavior of avalanches universally emerges.
These phenomena are studied in the context of self-organized criticality~\cite{sethna2001crackling}.
The steady state is automatically tuned on the critical point by the balance between external driving forces and their response.

Plastic event in the shear flow of amorphous solids is a representative example of such phenomena. 
Molecular dynamics (MD) simulations have played a key role to detect the avalanche criticality and to provide microscopic insights~\cite{maloney2006amorphous, oyama2021unified}. 
When amorphous solids are sheared, a rearrangement event first occurs locally, which gives rise to a stress fluctuation in the surrounding medium. 
This fluctuation sometimes triggers other local rearrangements, which lead to the cascade of rearrangements resulting in a rarely system-spanning relaxation event.
The sizes of these events, measured by the stress drop or the total squared displacements (TSD) of particles, follow a power-law distribution, which is a hallmark of critical avalanche dynamics. 
This view is further supported by the studies of the elastoplastic model (EPM), a coarse-grained model of amorphous solids, where the system is simplified into an assembly of mesoscopic elastoplastic blocks~\cite{nicolas2018deformation,talamali2011avalanches}. 
Interestingly, sheared amorphous solids exhibit a scale-free behavior of avalanches even not in the steady flow state. 
Both MD and EPM show that plastic events follow a power-law distribution even when a small shear strain is applied to quiescent amorphous solids, although the critical exponent is different from the one in the steady state~\cite{karmakar2010statistical, lin2016mean}. 
This characteristic criticality has been discussed in terms of the marginal stability of amorphous solids. 

The dynamics of supercooled liquids and glasses are known to exhibit intermittency even without shear~\cite{berthier2011theoretical}. 
Then a fundamental question naturally arises: whether avalanche criticality emerges in such glasses driven solely by thermal fluctuation. 
Recently, the thermal EPM was proposed as a coarse-grained model of unsheared supercooled liquids, where the dynamics is driven solely by thermal fluctuation~\cite{ozawa2023elasticity, tahaei2023scaling}. 
It was revealed that dynamically correlated regions emerge by stress relaxation triggered by thermal fluctuation and the dynamic facilitation by elastic fields~\cite{ozawa2023elasticity}.
In the thermal EPM, a certain avalanche dynamics was observed and the scaling relation between the avalanche criticality and the dynamical correlation length at finite temperatures was discussed~\cite{tahaei2023scaling}.
However, from a microscopic viewpoint, avalanche criticality has not been identified in experimental systems nor particle models driven only by thermal fluctuation.
Some of the experimental works and MD simulations reported intermittent or ``avalanche-like'' dynamics in aging~\cite{buisson2003intermittency, el2010subdiffusion, yanagishima2017common}, crystallization~\cite{sanz2014avalanches, yanagishima2017common}, phase-separation~\cite{testard2014intermittent}, and stationary dynamics in glasses~\cite{mizuno2020intermittent}. 
However, these studies do not quantify the probability distribution of the event sizes, presumably because huge numerical efforts are required. 
As a result, the above-mentioned question remains unanswered yet. 

This letter aims to answer this question. 
We use MD simulations to investigate the avalanche criticality in the dynamics of quiescent glasses. 
We prepare quiescent glasses by rapid quenching from high-temperature liquids and apply small thermal fluctuations to this model. 
This is the simplest setting of the aging dynamics.
We monitor the particle rearrangement events using the technique of the potential energy landscape analysis. 
We demonstrate that the rearrangement events consist of several local rearrangements linked by the elastic field.
The size of the rearrangement events follows a power-law distribution and finite-size scaling analysis shows that a scaling assumption is well satisfied.
These results suggest that the particle rearrangement driven by thermal fluctuation is a critical avalanche dynamics.
We then explain why the criticality emerges in our setting from the viewpoint of elastoplasticity.
Our study founds a firm ground to investigate the relation between intermittent dynamics and avalanche criticality in supercooled liquids and glasses. 

\textit{Model}.---We consider a 50:50 mixture of two types of soft sphere particles in two dimensions.
Type A and B particles have equal mass $m$ but different particle sizes to avoid crystallization with the diameter ratio $\sigma_B/\sigma_A=1.4$.
The interaction potential is an inverse-power-law potential given by $v(r_{ij})=\varepsilon\lbrack\left({\sigma_{ij}}/{r_{ij}}\right)^{12}-\left({\sigma_{ij}}/{r_{c,ij}}\right)^{12}+C\left({(r_{ij}-r_{c,ij})}/{\sigma_{ij}}\right)+D\left({(r_{ij}-r_{c,ij})}/{\sigma_{ij}}\right)^2\rbrack\theta(r_{c,ij}-r_{ij})$,
where $\varepsilon$ is an energy scale; $r_{ij}=\left|\bm{r}_i-\bm{r}_j\right|$, with $\bm{r}_i$ being the position of particle $i$; and $\sigma_{ij}\equiv\left(\sigma_i+\sigma_j\right)/2$, where $\sigma_{i}$ represents the diameter of particle $i$.
The step function $\theta(r_{c,ij}-r_{ij})$ represents the cutoff of the potential with the cutoff length $r_{c,ij}\equiv1.5\sigma_{ij}$.
The coefficients $C$ and $D$ force the continuity of the first and second derivatives at $r_{c,ij}$.
To approximate a bulk system, we use the periodic boundary condition in the two-dimensional system (volume $V$).
Mass, length, time, and temperature are measured in units of $m$, $\sigma_A$, $\sigma_A\sqrt{m/\varepsilon}$ and $\varepsilon/k_{\mathrm{B}}$ respectively, where $k_{\mathrm{B}}$ is the Boltzmann constant.
The number of particles $N$ is $324,~1024,~3136,~10000$, and $32400$.
The number density $\rho = N/V$ is fixed to $0.8$. 

We first perform MD simulations of the model with the Nos\'{e}--Hoover thermostat at the temperature $T=10$ to obtain equilibrium configurations. 
We then apply the optimization method FIRE2.0 to these configurations to obtain the inherent structures (ISs, nearest neighbor energy minimum), which are the initial states of our production runs~\cite{guenole2020assessment}. 
As the production runs, we solve the underdamped Langevin equation starting from the ISs for the time duration $t_{\rm sim}$ at the temperature $T$. 
The dissipation is modeled by the Stokes drag force $\bm{f}_i^{\mathrm{drag}}=-\gamma \dot{\bm{r}}_i$ with the damping coefficient $\gamma=1$. 
The time step of the MD simulation is 0.001.
We mainly present the results with $T=0.1$ and $t_\mathrm{sim} = 100$ unless otherwise noted. 
Since $T=0.1$ is much below the computational glass transition temperature of the model, our production run corresponds to the aging dynamics. 
Note that, to confirm the generality of our results, we additionally performed the production runs with the Nos\'{e}--Hoover thermostat and obtained similar results (see supplemental material(SM)~VI).

\begin{figure}
\centering
\includegraphics[width=\columnwidth]{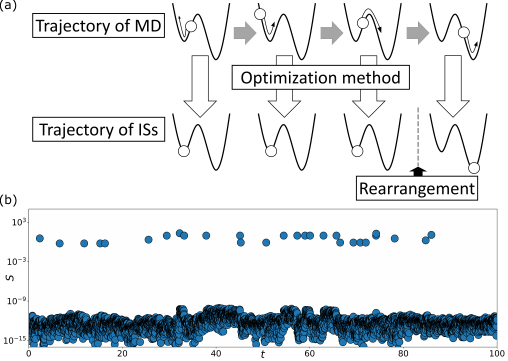}
\caption{(a) The schematic figure of our analysis. We first obtain MD trajectories and calculate ISs of them by the optimization method.
We detect the particle rearrangements as the change of the ISs.
(b) TSD $S(t)$ of single tajectory with $N=3136$.
}
\label{fig:Analysis}
\end{figure}

\begin{figure*}
\centering
\includegraphics[width=\textwidth]{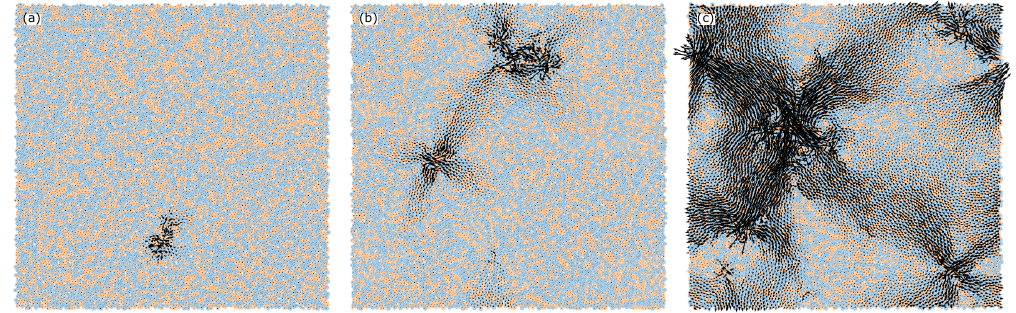}
\caption{Displacement fields of rearrangement events with $N=10000$.
Each figure corresponds to the rearrangement event with (a) $S=1$, (b) $S=10$, and (c) $S=100$.
Orange and blue circles represent the particles of type A and type B, respectively.
The arrows that represent particle displacement are scaled for visibility.
}
\label{fig:Displacement}
\end{figure*}

\textit{Detection of rearrangements}.---By the protocol outlined above, we obtain the time series of the configurations during the aging dynamics. 
To detect rearrangements, we performed the following potential energy landscape analysis. 
First, we discretize the time series with a time interval $\delta t=0.02$, hence the series consists of $t_\mathrm{sim}/ \delta t = 5000$ successive configurations. 
Next, we apply FIRE2.0 to these configurations to obtain the corresponding ISs~\cite{guenole2020assessment}.
This analysis allows us to remove the effect of vibrations and focus only on the rearrangements, which correspond to transitions between the ISs (see Fig.~\ref{fig:Analysis}(a)).

We now define the size of the rearrangements. 
Hereafter, $\bm{r}_{i,\mathrm{IS}}(t)$ denote the position of the particle $i$ in the IS at time $t$. 
In studies of sheared amorphous solids, the stress or energy drop is often used as an indicator of the size of the rearrangement.
However, these quantities are not appropriate for our setting because these quantities can be either positive or negative due to the isotropic and thermal nature of the system. 
Therefore, we define the size of rearrangements as the TSD of the particles between the ISs: $S(t)= \sum_i (\bm{r}_{i,\mathrm{IS}}(t)-\bm{r}_{i,\mathrm{IS}}(t+\delta t))^2$.
Note that the TSD also exhibits the criticality in sheared amorphous solids~\cite{oyama2023shear}. 

Figure~\ref{fig:Analysis}(b) shows TSD $S(t)$ for a single production run with $N=3136$.
There are two kinds of events: events with $S(t)<10^{-8}$ and events with $S(t)>10^{-2}$. 
Considering the limitation of the double-precision arithmetic, $S(t)<10^{-8}$ indicates that no rearrangement has occurred. 
On the other hand, $S(t)>10^{-2}$ indicates that the ISs at $t$ and $t+\delta t$ are different and the rearrangement has occurred.
We define the events with $S(t)>S_\mathrm{th}=10^{-6}$ as rearrangement events.
We confirmed that the threshold $S_\mathrm{th}$ within $10^{-8} \le S_\mathrm{th} \le 10^{-2}$ returns the completely same set of the rearrangement events.
We note that $S(t)$ depends on $\delta t$ because too large $\delta t$ may mix up multiple transitions between ISs into a single event. 
We confirmed that $\delta t=0.02$ and $0.05$ give almost the same results, suggesting that the impact of the multiple transitions is negligible for $\delta t=0.02$ (see SM~I). 

\textit{Displacement in real space}.---We first show the real-space properties of the rearrangement events.
Figure~\ref{fig:Displacement} shows the displacements of particles in three different-sized rearrangement events. 
Only one local region is in large motion for the event with $S=1$, while two local rearrangements appear for the event with $S=10$. 
For the event with $S=100$, there are multiple local rearrangements to form a system-spanning event. 
We emphasize that this system-spanning rearrangement has occurred as a single transition between two ISs, as in the case of sheared amorphous solids.  

From the visual inspection, we find that the local rearrangements are coupled by the elastic field.
The reason is that the displacement fields around the local rearrangement regions have a clear quadrupolar character and multiple rearrangement regions are linked by this field. 
This is a signature of the Eshelby field often observed in two-dimensional elastic bodies and is similar to the displacement fields in sheared amorphous solids~\cite{maloney2006amorphous}.

\begin{figure*}
\centering
\includegraphics[width=\textwidth]{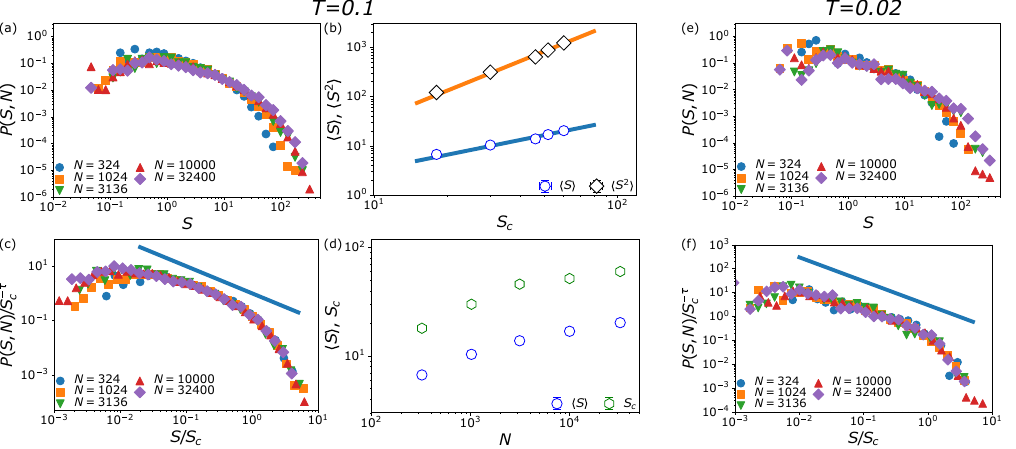}
\caption{(a) Probability distributions $P(S,N)$ of the size $S$ of rearrangement events with different system size $N$.
(b) Dependence of $\langle S\rangle$ and $\langle S^2 \rangle$ on the cutoff $S_c=\langle S^2 \rangle/\langle S\rangle$.
Orange and blue lines are proportional to $S_c^2$ and $S_c^1$ respectively, which correspond to the exponent $\tau=1$ of the distribution.
(c) Scaled probability distribution. The curves collapse on the same curve. Blue line represents the slope of $S^{-\tau}\sim S^{-1}$.
(d) System size dependence of $\langle S\rangle$ and $\langle S^2 \rangle$.
(e) Probability distributions $P(S,N)$ of the size $S$ of rearrangement events with different system size $N$ at $T=0.02$.
(f) Scaled probability distribution at $T=0.02$. The curves collapse on the same curve. Blue line represents the slope of $S^{-\tau}\sim S^{-1}$.
The error bars in (b) and (d) represent bootstrap standard errors (see SM~II).
}
\label{fig:Scaling}
\end{figure*}

\textit{Statistical property}.---We now analyze the statistical properties of the rearrangements. 
To this end, we perform a large set of simulations: $1000$ independent trajectories for $N=324$ and $1024$, $100$ trajectories for $N=3136$ and $10000$, and $10$ trajectories for $N=32400$.
We then calculate the probability distribution $P(S,N)$ of the TSD $S$ of the rearrangements for various $N$.
Figure~\ref{fig:Scaling}(a) shows that $P(S,N)$ has a power-law region and the region extends with increasing $N$. 
To ascertain whether the power-law distribution originates from a critical phenomenon, we perform a scaling analysis.
We assume the standard scaling relation 
\begin{eqnarray}
P(S,N)  &\sim&  S^{-\tau}f(S/S_c)=S_c^{-\tau}g(S/S_c),\label{eq:scaling_assumption}
\end{eqnarray}
where $S_c$ is an $N$-dependent cutoff size and $f(x)$ and $g(x) \equiv x^{-\tau}f(x)$ are scaling functions.
Then from Eq.~\eqref{eq:scaling_assumption}, the first and second moments $\langle S\rangle=\int_0^\infty SP(S,N)dS$ and $\langle S^2\rangle=\int_0^\infty S^2P(S,N)dS$ follow 
\begin{eqnarray}
\langle S\rangle \sim S_c^{2-\tau},~\langle S^2\rangle \sim S_c^{3-\tau} \label{eq:moment_and_cutoff}
\end{eqnarray}
for $1<\tau<2$, where $\langle \rangle$ denotes the average on the distribution.
Guided by Eq.~\eqref{eq:moment_and_cutoff}, we define the cutoff size as $S_c\equiv\langle S^2\rangle/\langle S\rangle$, omitting unimportant numerical constants. 
We numerically measured $\langle S\rangle$ and $\langle S^2 \rangle$, and plotted them against $S_c = \langle S^2\rangle/\langle S\rangle$ in Fig.~\ref{fig:Scaling}(b). 
Clearly, the numerical data follows a power-law relationship, which is consistent with Eq.~(\ref{eq:moment_and_cutoff}).
From this plot, we estimated the exponent as $\tau\simeq1$ (see SM~III). 
Next, to directly verify the scaling relation Eq.~(\ref{eq:scaling_assumption}), we plotted $P(S,N)/S_c^{-\tau}$ against $S/S_c$ with the exponent $\tau=1$ in Fig.~\ref{fig:Scaling}(c).
The distributions with different system sizes collapse well on a single curve.
We also find that the solid line in Fig~\ref{fig:Scaling}(c), which indicates the power-law relation $P(S,N)  \propto  S^{-\tau}$ with $\tau=1$, describes well our numerical data. 
These results suggest that the scaling hypothesis Eq.~(\ref{eq:scaling_assumption}) works well and that a critical phenomenon results in the power-law distribution with $\tau=1$. 
We also analyze the trajectories with lower temperature $T=0.02$ in the same way and found that the scaling hypothesis with $\tau\simeq1$ still works (see Figs.~\ref{fig:Scaling}(e)(f) and SM~\ref{SM:lowT}).
Moreover, the same scaling relation is confirmed in the situation with shorter simulation time $t_\mathrm{sim}$, with a different thermostat, and with a different definition of event size~\cite{nishikawa2022relaxation}, which implies the generality of our findings (see SM~IV, VI, and VII).

We note that in our scaling discussion, the system size dependence of the cutoff $S_c$ does not necessarily follow a power law.
In actual fact, we observed that $\langle S \rangle$ and $S_c$ increase with $N$ in a power-law for smaller $N$, but the increase becomes milder for larger $N$ (see Fig.~\ref{fig:Scaling}(d)). 
Though the ultimate fate of the cutoff size $S_c$ at large $N$ is an important issue, the required computational cost is beyond our reach for the moment. 
We thus leave this for future work~\footnote{ 
In larger systems, the rearrangements occur more frequently, and then we need to use smaller time intervals $\delta t$ to isolate individual rearrangement events. 
This requires huge numerical costs. }.

\textit{Discussion}.--Our result is the first observation of the avalanche criticality of rearrangements in glasses driven solely by thermal fluctuation. 
Here, we discuss this result from the elastoplastic viewpoint. 
From the elastoplastic viewpoint, the avalanche dynamics in purely thermal glasses can emerge via the following mechanism~\cite{tahaei2023scaling}. 
Due to the thermal fluctuation, the system sometimes undergoes a crossing of an energy barrier, which corresponds to a local rearrangement of particles. 
This rearrangement causes the elastic response of the surrounding medium, which results in the fluctuations of the energy barriers of the surrounding medium. 
In some cases, this fluctuation leads to the subsequent local rearrangements, which constitute the avalanche. 
This elastoplastic view is broadly consistent with our observations of the rearrangement events: thermal fluctuation causes transitions between ISs and the elastic field links multiple local rearrangements. 

\begin{figure}
\centering
\includegraphics[width=\columnwidth]{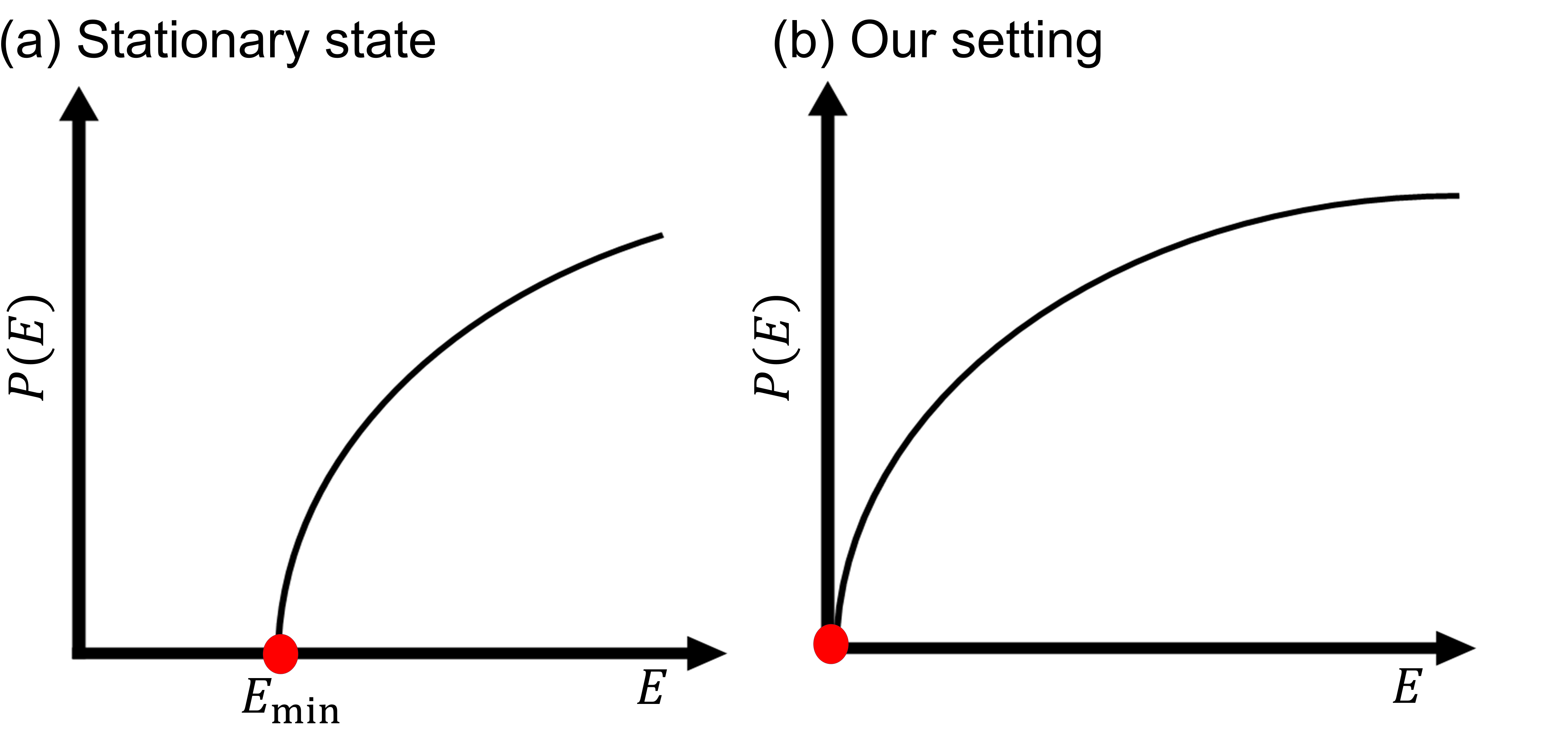}
\caption{Rough sketch of the distribution of activation energy with (a) stationary condition and (b) the state just after fast quench (our setting).
}
\label{fig:Schematic}
\end{figure}

For a deeper understanding, we use recent findings on the thermal EPM. 
The thermal EPM is a model of an assembly of mesoscopic blocks agitated solely by thermal fluctuation~\cite{ozawa2023elasticity}. 
Each block undergoes a plastic event when a crossing of the energy barrier occurs. 
Though the aging dynamics is not studied within this model, its stationary state was recently studied intensively~\cite{ozawa2023elasticity,tahaei2023scaling}. 
At low enough temperature, the model reaches the stationary state in which the probability distribution of the energy barrier becomes gapped (see Fig.~\ref{fig:Schematic}(a)). 
The edge value $E_\mathrm{min}$ of the energy barrier dominates the relaxation time of the system as $\tau_\alpha \sim \exp(E_\mathrm{min}/k_{\mathrm{B}}T)$. 
Moreover, the critical avalanche dynamics could be observed in this setting if the plastic events that occur within the time duration $\delta t \sim \tau_\alpha$ are regarded as a single plastic event~\cite{tahaei2023scaling}. 

We can use these insights to interpret our results. 
We found the avalanche criticality in glasses immediately after a quench. 
Importantly, previous MD simulations and the EPM studies established that the energy barrier for the rearrangement becomes vanishingly small in such systems~\cite{lin2016mean, karmakar2010statistical}. 
This means that the edge $E_\mathrm{min}$ of the activation energy distribution is equal to zero, and thus the activation time is the microscopic time scale (see Fig. ~\ref{fig:Schematic}(b)). 
This observation suggests that the avalanche criticality can emerge in our setting with $\delta t \to 0$, which is exactly what we found in our numerical simulations. 
Moreover, this discussion suggests that the avalanche dynamics disappears if the aging dynamics after finite waiting time is focused.
Indeed, this is consistent with the known results: intermittent rearrangements occur during the aging after finite waiting time but these events are not system-spanning nor critical~\cite{el2010subdiffusion, mizuno2020intermittent}. 

Quantitatively, the exponent $\tau \simeq 1$ found in this work is significantly smaller than the ones observed in the stationary state of the thermal EPM ($\tau \simeq 1.3$~\cite{tahaei2023scaling}). 
The steady shear flow of amorphous solids is also characterized by larger exponents: $\tau \simeq 1.36$ in the two-dimensional EPM~\cite{lin2014scaling} and $\tau \simeq 1.49$ in the MD simulations with an athermal quasistatic protocol~\cite{oyama2021unified}, though the differences in dynamics and protocols can affect the value of $\tau$~\cite{papanikolaou2016shearing, jagla2015avalanche}.
On the other hand, $\tau \simeq 1$ was predicted by random energy model~\cite{franz2017mean} and reported for the elastic regime of the sheared amorphous solids~\cite{shang2020elastic}, where the system is close to the glasses just after quench. 
These observations hint at a new characterization of the difference between the equilibrium/stationary state dynamics and the aging dynamics of glasses: the avalanche critical exponent. 

\textit{Conclusion and future perspective}.--We investigate rearrangement events in quiescent glasses just after a quench by MD simulations.
The rearrangement events exhibit displacement fields consisting of local rearrangements interacting via elastic fields.
The size distribution of the rearrangements shows a power-law behavior and the finite-size scaling analysis can collapse the distribution for different system sizes.
Such statistical properties indicate that particle rearrangements can be understood as a critical phenomenon analogous to the fluidization of amorphous solids.
We can explain why the avalanche criticality was observed in our setting from the viewpoint of elastoplasticity.

Our study suggests that the aging dynamics in glasses can be understood by considering the glassy systems as “solids that flow” proposed in the studies of relaxation of supercooled liquids~\cite{dyre2006colloquium, lemaitre2014structural}.
The low-temperature relaxation of supercooled liquids has been studied by considering them as almost solids whose structure relaxes by rare local particle rearrangements and the resulting elastic dynamic facilitation.
Our study reveals that elastic dynamic facilitation exists even in aging glasses.
Although the aging has been studied by mean-field~\cite{cugliandolo1993analytical, cugliandolo1994out} and trap models~\cite{odagaki1990stochastic, monthus1996models}, a finite-dimensional theoretical approach describing intermittent rearrangements has been lacking.
Studying thermal EPMs, which model the concept of “ solids that flow,” can be one useful approach to understanding aging in real structural glasses.
The waiting time and maturity dependence of rearrangement should be investigated in this way and compared with experimental studies. 

Our analysis from the viewpoint of the potential energy landscape is versatile to measure rearrangements in a variety of situations. 
For example, this method is applicable to the equilibrium relaxation dynamics of supercooled liquids, and this is a promising way to detect the avalanche criticality in the equilibrium dynamics of supercooled liquids within the MD simulations. 


\begin{acknowledgments}
We thank Norihiro Oyama for useful discussions.
This work is supported by JSPS KAKENHI (Grant Numbers 20H01868, 22K03543, 23H04495, 24H00192, and 24KJ0589). 
\end{acknowledgments}

\bibliography{apssamp}

\end{document}


\newcommand{\beginsupplement}{
        \setcounter{table}{0}
        \renewcommand{\thetable}{S\arabic{table}}
        \setcounter{figure}{0}
        \renewcommand{\thefigure}{S\arabic{figure}}
        \setcounter{equation}{0}
        \renewcommand{\theequation}{S\arabic{equation}}
        \setcounter{section}{0}
     }

\preprint{APS/123-QED}
\title{Supplemental Material for\\ ``Avalanche criticality emerges by thermal fluctuation in a quiescent glass''}

\author{Yuki Takaha}
 \email{yukitakaha@g.ecc.u-tokyo.ac.jp}
 \affiliation{Graduate School of Arts and Sciences, The University of Tokyo, 3-8-1 Komaba, Tokyo 153-8902, Japan}
\author{Hideyuki Mizuno}
 \affiliation{Graduate School of Arts and Sciences, The University of Tokyo, 3-8-1 Komaba, Tokyo 153-8902, Japan}
\author{Atsushi Ikeda}
 \affiliation{Graduate School of Arts and Sciences, The University of Tokyo, 3-8-1 Komaba, Tokyo 153-8902, Japan}
 \affiliation{Research Center for Complex Systems Biology, Universal Biology Institute, The University of Tokyo, Komaba, Tokyo 153-8902, Japan}
 
\maketitle
\beginsupplement

\section{Dependence on $\delta t$}\label{SM:deltat}

The definition of $S$ depends on the time interval $\delta t$ with which we observe the rearrangements.
With larger $\delta t$ our analysis recognizes multiple transitions between ISs as a single event, so there are fewer apparent total events, a smaller proportion of small $S$ events, and a larger proportion of large $S$ events.
Due to this effect, a larger $\delta t$ makes $\langle S \rangle$ and $S_c$ larger.
However, if we reduce $\delta t$ sufficiently small, we can decompose the transitions between ISs one by one, and the physical quantities are expected to converge to a certain value.

\begin{figure}
\centering
\includegraphics[width=0.7\columnwidth]{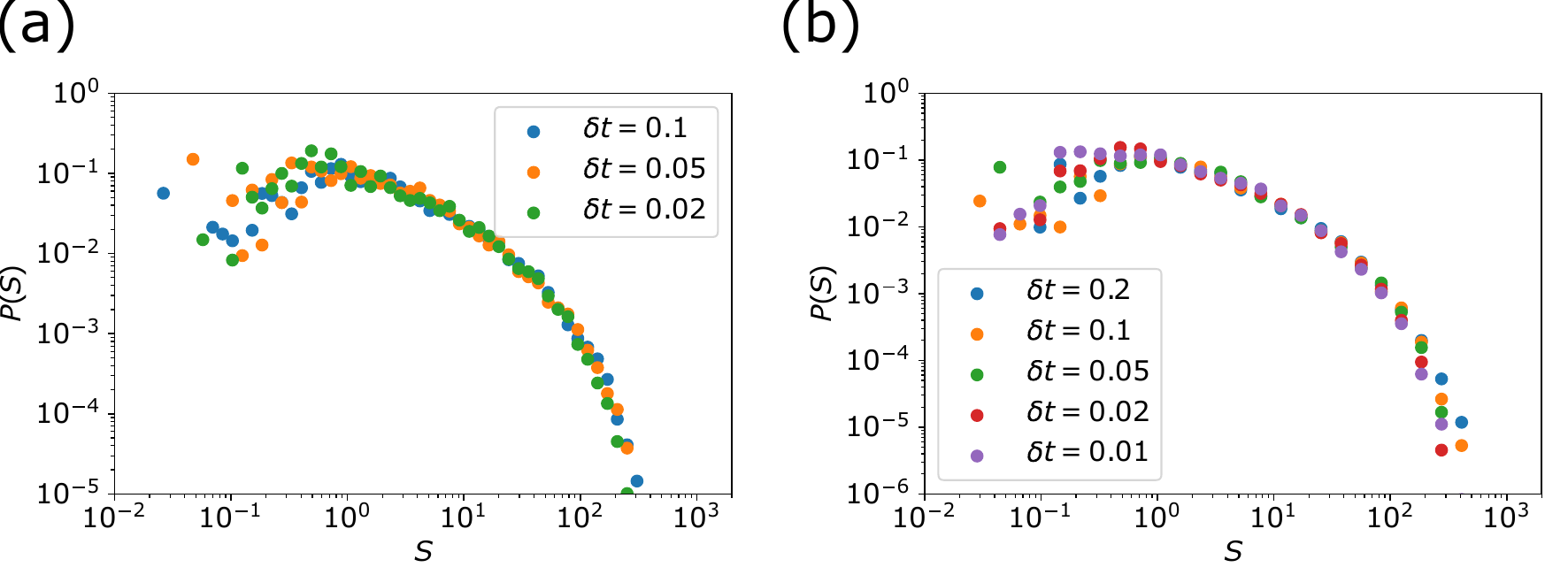}
\caption{(a) Probability distributions $P(S)$ of the size of rearrangement events with $N=324$ and various $\delta t$.
(b) Probability distributions $P(S)$ of the size of rearrangement events with $N=32400$ and various $\delta t$.
}

\label{fig:Supplement_deltat}
\end{figure}

Figures~\ref{fig:Supplement_deltat}(a) and (b) shows the distribution $P(S)$ of the size of rearrangements with different $\delta t$ for $N=324$ and $N=32400$.
For $N=324$ and $N=32400$, almost no $\delta t$ dependence is observed in the distribution, and $\delta t=0.02$ is considered small enough for small systems.

\section{Evaluation of errors of $\langle S \rangle$, $\langle S^2 \rangle$ and $\langle S_c \rangle$}\label{SM:bootstrap}

Since the moments and cutoffs of the size of rearrangements are themselves statistic quantities, we evaluate the errors by the bootstrap method.
First, we prepare a set of the size $S$ of the observed rearrangement events for each system size.
We define the number $N_\mathrm{event}$ of rearrangement events.
Next, we randomly resample the values from these sets $N_\mathrm{event}$ times, allowing for duplicates.
Same resampling is performed $N_{\mathrm{bootstrap}}$ times to obtain $N_{\mathrm{bootstrap}}$ new sets of $S$.
Finally, we compute $\langle S\rangle,~\langle S^2\rangle$ and $S_c=\langle S^2\rangle/\langle S\rangle$ on each resampled set, and get the distributions of them.
From these distributions, standard deviations are computed and these were used as the error for each value.
The number of bootstrap samples is set to $N_{\mathrm{bootstrap}}=200$.

\section{The value of the exponent $\tau$}\label{SM:tau}
\begin{figure}
\centering
\includegraphics[width=0.7\columnwidth]{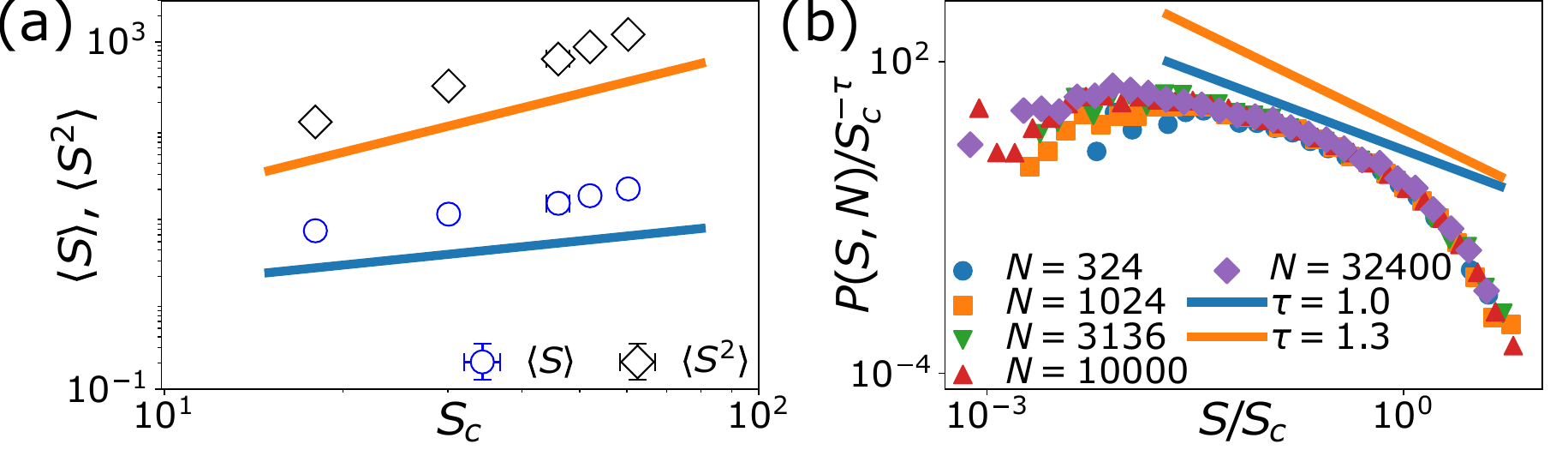}
\caption{(a) Dependence of $\langle S\rangle$ and $\langle S^2 \rangle$ on the cutoff $S_c=\langle S^2 \rangle/\langle S\rangle$.
Orange and blue line are proportional to $S_c^{1.7}$ and $S_c^{0.7}$ respectively, which correspond to the exponent $\tau=1.3$.
(b) Scaled probability distribution with $\tau=1.3$ and lines which represent the power law dependence with $\tau=1.0$ and $1.3$. The power law region of the distribution is with the exponent smaller than $1.3$.
}

\label{fig:Supplement_tau}
\end{figure}

Analysis using the value of the exponent $\tau\simeq1.3$, observed in the thermal EPM, does not explain the exponent in the power-law region of the probability distribution of rearrangement size, although it may explain the cutoff dependence of the moments and collapse the distribution with different system size.
Figure~\ref{fig:Supplement_tau}(a) shows the cutoff dependence of the moments and a straight line corresponding to the exponent $\tau=1.3$.
The exponent appears to be in good agreement.
Figure~\ref{fig:Supplement_tau}(b) is a collapsed distribution with $\tau=1.3$.
While the distributions for different system sizes collapse well, the exponent in the power-law region of the distribution is smaller than 1.3.
From these observations, we conclude that the exponent is $\tau \simeq1$.

\section{Statistical analysis with shorter observation time}\label{SM:shorttime}

\begin{figure}
\centering
\includegraphics[width=\columnwidth]{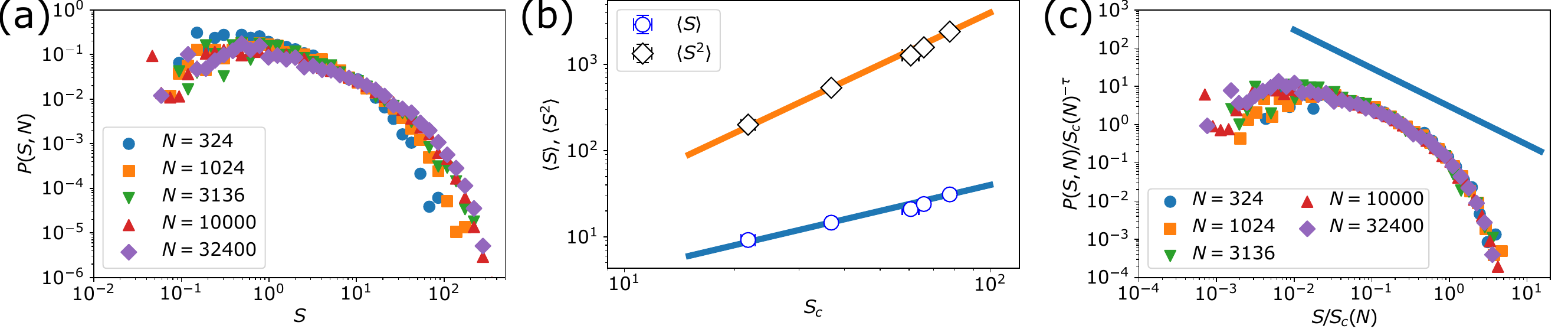}
\caption{(a) Probability distributions $P(S,N)$ of the size of rearrangement events over $0\le t<20$.
(b) Dependence of $\langle S\rangle$ and $\langle S^2 \rangle$ on the cutoff $S_c=\langle S^2 \rangle/\langle S\rangle$.
Orange and blue lines are proportional to $S_c^2$ and $S_c^1$ respectively, which correspond to the exponent $\tau=1$ of the distribution.
The errorbars represent the bootstrap standard errors(see SM.~\ref{SM:bootstrap}).
(c) Scaled probability distribution. The curves collapse on the same curve. Blue line represents the slope of $S^{-\tau}\sim S^{-1}$.
}

\label{fig:Supplement_shorttime}
\end{figure}

The statistical results obtained in the main text are also obtained for shorter-time observations.
We computed the distribution only for rearrangements observed at $0\le t<20$.
Figure~\ref{fig:Supplement_shorttime}(a) shows the distribution $P(S,N)$ of the size of the rearrangements observed for $0\le t<20$.
The moments $\langle S\rangle,~\langle S^2\rangle$ and the cutoff $S_c=\langle S^2\rangle/\langle S\rangle$ satisfies the scaling relation, whose exponent is $\tau\simeq 1$ (see Fig.~\ref{fig:Supplement_shorttime}(b)).
Figure~\ref{fig:Supplement_shorttime}(c) shows that finite size scaling with the cutoff $S_c$ confirms that the distribution collapses on the same curve and has a power-law region corresponding to $S^{-\tau}\sim S^{-1}$.

\section{Statistical analysis with lower temperature}\label{SM:lowT}

\begin{figure}
\centering
\includegraphics[width=\columnwidth]{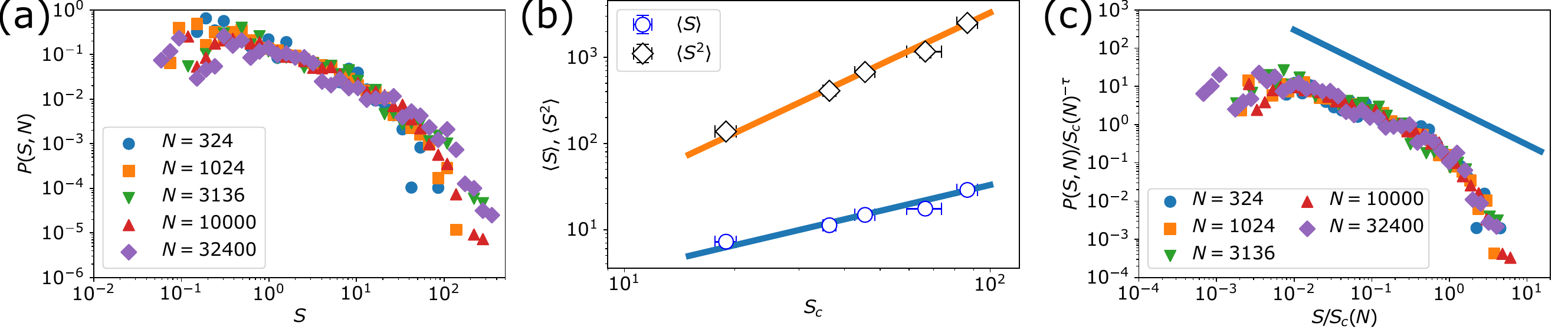}
\caption{(a) Probability distributions $P(S,N)$ of the size of rearrangement events with $T=0.02$.
(b) Dependence of $\langle S\rangle$ and $\langle S^2 \rangle$ on the cutoff $S_c=\langle S^2 \rangle/\langle S\rangle$.
Orange and blue lines are proportional to $S_c^2$ and $S_c^1$ respectively, which correspond to the exponent $\tau=1$ of the distribution.
The errorbars represent the bootstrap standard errors(see SM.~\ref{SM:bootstrap}).
(c) Scaled probability distribution. The curves collapse on the same curve. Blue line represents the slope of $S^{-\tau}\sim S^{-1}$.
}

\label{fig:Supplement_lowT}
\end{figure}

We analyze the MD data at a lower temperature in the same way.
We perform the same MD simulations at $T=0.02$ with the same initial configurations as in the main text.
We detect the particle rearrangements with time interval $\delta t = 0.1$.
We obtain the distribution of rearrangements with exactly the same analysis.
Figure~\ref{fig:Supplement_lowT}(a) shows the distribution $P(S,N)$ of the size of the rearrangements at $T=0.02$.
The distribution is rough due to the fewer rearrangements at lower temperatures.
We confirm the power-law relation between $\langle S\rangle,~\langle S^2\rangle$ and $S_c=\langle S^2\rangle/\langle S\rangle$ calculated from these distributions, which corresponds to $\tau\simeq 1$(see Fig.~\ref{fig:Supplement_lowT}).
Figure ~\ref{fig:Supplement_lowT}(c) shows that finite size scaling confirms that the distribution collapses on the same curve and has a power-law region corresponding to $S^{-\tau}\sim S^{-1}$.

\section{Statistical analysis with Nos\'{e}--Hoover thermostat}\label{SM:Nose}
\begin{figure}
\centering
\includegraphics[width=\columnwidth]{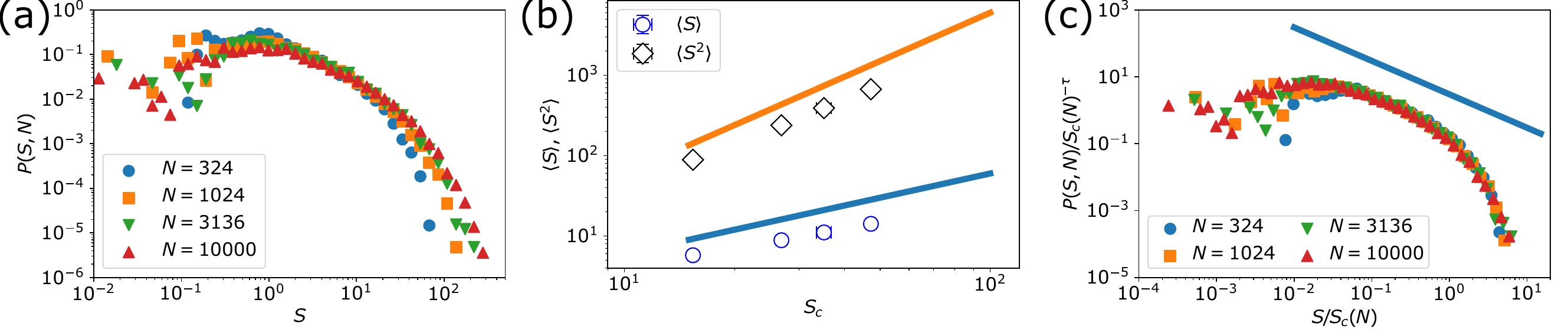}
\caption{(a) Probability distributions $P(S,N)$ of the size of rearrangement events in the dynamics with Nos\'{e}--Hoover thermostat.
(b) Dependence of $\langle S\rangle$ and $\langle S^2 \rangle$ on the cutoff $S_c=\langle S^2 \rangle/\langle S\rangle$.
Orange and blue lines are proportional to $S_c^2$ and $S_c^1$ respectively, which correspond to the exponent $\tau=1$ of the distribution.
The errorbars represent the bootstrap standard errors(see SM.~\ref{SM:bootstrap}).
(c) Scaled probability distribution. The curves collapse on the same curve. Blue line represents the slope of $S^{-\tau}\sim S^{-1}$.
}
\label{fig:Supplement_nose}
\end{figure}

The scaling relation is also confirmed for the dynamics with the Nos\'{e}--Hoover thermostat.
We perform MD simulations with the Nos\'{e}--Hoover thermostat at $T=0.1$ with the same initial configurations as in the main text.
We detect the particle rearrangements with time interval $\delta t = 0.1$, which is almost equal to the time scale of the vibration of the thermostat.
We obtained the distribution of rearrangements with the same analysis.
Figure~\ref{fig:Supplement_nose}(a) shows the distribution $P(S,N)$ of the size of the rearrangements with the Nos\'{e}--Hoover thermostat.
There is a power-law relation between the moments $\langle S\rangle,~\langle S^2\rangle$ and the cutoff $S_c=\langle S^2\rangle/\langle S\rangle$ calculated from these distributions, which corresponds to $\tau\simeq 1$(see Fig.~\ref{fig:Supplement_nose}).
Figure~\ref{fig:Supplement_nose}(c) shows that finite size scaling confirms that the distribution collapses on the same curve and has a power-law region corresponding to $S^{-\tau}\sim S^{-1}$.

\begin{figure}
\centering
\includegraphics[width=\columnwidth]{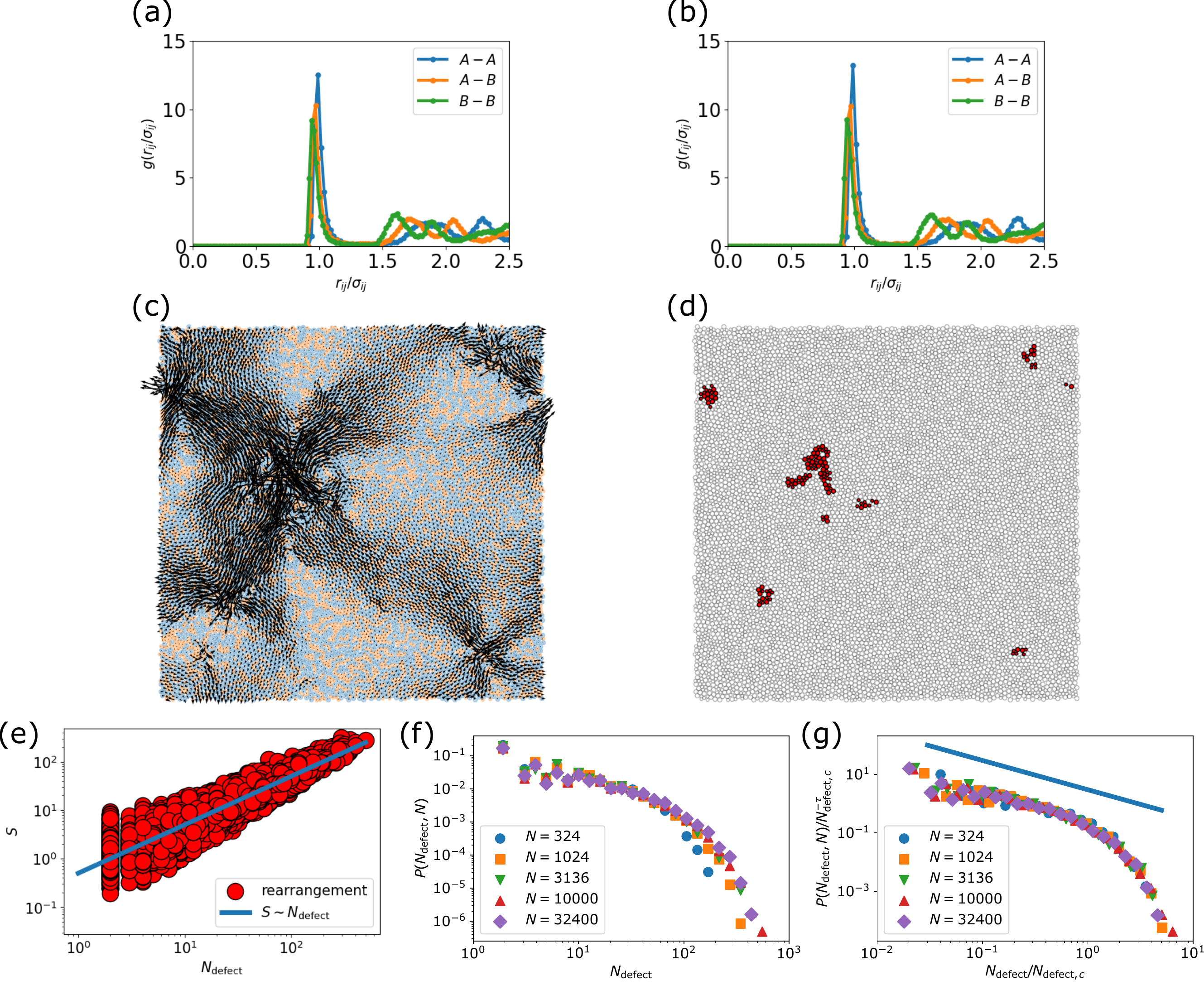}
\caption{(a)(b) Scaled radial distribution function $g(r_{ij}/\sigma_{ij})$ (a) at $t=0$ and (b) at $t=100$ with $N=10000$.
(c) Displacement field and (d) defect field of the same event.
(e) The scatter plot of relation between TSD $S$ and the number of defect $N_{\mathrm{defect}}$ of particle rearrangement events with $N=10000$. The blue line indicates the linear relation between $S$ and $N_{\mathrm{defect}}$.
(f) Probability distributions $P(N_\mathrm{defect},N)$ of the number of defects.
(g) Probability distribution scaled with the cutoff $N_{\mathrm{defect},c}$. The curves collapse on the same curve. Blue line represents the slope of $N_\mathrm{defect}^{-\tau}\sim N_\mathrm{defect}^{-1}$.
}
\label{fig:Supplement_defect}
\end{figure}

\section{Definition of avalanche size by the number of defects}\label{SM:defect}
We analyze the MD trajectory with another definition as the size of particle rearrangements based on the defect field~\cite{nishikawa2022relaxation}.
In this definition, we detect the local plastic events based on breakages and insertions of bonds between the particles from time $t$ to $t+\delta t$.
Thus, the number of defects is not affected by elastic deformation around local plastic events, which is accumulated in the total square displacements $S$, the definition in the main text. 

Before we define the bond-breakage and bond-insertion process, we monitor the radial distribution functions of ISs.
We calculate scaled radial distribution functions $g(r_{ij}/\sigma_{ij})$ between small particles ($A-A$), large particles($B-B$), and a small particle and a large particle ($A-B$) as a function of a normalized distance $r_{ij}/\sigma_{ij}$ at $t=0$ and $t=100$ for a single sample with $N=10000$ (see Figs.~\ref{fig:Supplement_defect}(a) and (b)).
The radial distribution functions are not changed much within this observation time.
We confirm a nearly flat region ($r_{ij}/\sigma_{ij}\approx1.2-1.4$) between the first peak and the second peak.
We define the scaled minimum distance $x_{\mathrm{min}}$ to the neighbors by $x_{\mathrm{min}}=1.2\sigma_{ij}$ and the scaled maximum distance $x_{\mathrm{max}}$ by $x_{\mathrm{max}}=1.4\sigma_{ij}$.

We first analyze the bond-breakage process between time $t$ and $t+\delta t$.
We consider that there is a bond between particles $i$ and $j$ at time $t$ if the particles are located within $r_{ij}<x_{\mathrm{min}}$.
If the bond is $r_{ij}\ge x_\mathrm{max}$ at time $t+\delta t$, we regard the bond is broken.
We next analyze the bond-insertion process between time $t$ and $t+\delta t$.
We focus particles $i$ and $j$ such that $r_{ij}\ge x_{\mathrm{max}}$ is satisfied at time $t$ and consider there is no bond between them.
Then, we regard that the bond between the particles $i$ and $j$ is generated at time $t+\delta t$ if the particles become located within $r_{ij}<x_{\mathrm{min}}$.
We define the defect order parameter $\phi_i$ as follows: $\phi_i=0$ if there is no bond broken or inserted at time $t+\delta t$ between particle $i$ and any other particles, and $\phi_i=1$ if there are bonds broken or inserted at time $t+\delta t$ between particle $i$ and other particles.
We can confirm that the defect field indicates the particles participating in local plastic events (see Figs.~\ref{fig:Supplement_defect}(c) and (d)).
Finally, we define the number of defects $N_{\mathrm{defect}}$ by $N_{\mathrm{defect}}=\sum_i\phi_i$.
Figure~\ref{fig:Supplement_defect}(e) shows $S$ and $N_{\mathrm{defect}}$ of the same rearrangement events have a linear relation.

The choice of the definition does not affect the result and discussion in the main text.
Figure~\ref{fig:Supplement_defect}(f) shows the probability distributions $P(N_\mathrm{defect},N)$ of the number of defects.
We calculate the moments and the cutoff of the number of defects in the same way in the main text. 
The finite size scaling shows that the distributions collapse on the same curve (see Fig.~\ref{fig:Supplement_defect}(g)) and the scaling relation holds for $N_\mathrm{defect}$.
Especially, the exponent of the power-law region is also $\tau\simeq1$ for the probability distribution of $N_{\mathrm{defect}}$.

\bibliography{apssamp}